# Magnetic Quantum Tunneling: Insights from Simple Molecule-Based Magnets

Stephen Hill,[1] Saiti Datta,[1] Junjie Liu,[2] Ross Inglis,[3] Constantinos J. Milios,[3] Patrick L. Feng,[4] John J. Henderson,[5] Enrique del Barco,[5] Euan K. Brechin[3] and David N. Hendrickson[4]

[1] *NHMFL and Department of Physics, Florida State University, Tallahassee, FL 32310, USA*
[2] *Department of Physics, University of Florida, Gainesville, FL 32611, USA*
[3] *School of Chemistry, The University of Edinburgh, EH9 3JJ, U.K.*
[4] *Department of Chemistry and Biochemistry, UC San Diego, La Jolla, CA 92093, USA*
[5] *Department of Physics, University of Central Florida, Orlando, FL 32816, USA*

This perspectives article takes a broad view of the current understanding of magnetic bistability and magnetic quantum tunneling in single-molecule magnets (SMMs), focusing on three families of relatively simple, low-nuclearity transition metal clusters: spin $S = 4$ $Ni_4^{II}$, $Mn_3^{III}$ ($S = 2$ and 6) and $Mn_6^{III}$ ($S = 4$ and 12). The $Mn^{III}$ complexes are related by the fact that they contain triangular $Mn_3^{III}$ units in which the exchange may be switched from antiferromagnetic to ferromagnetic without significantly altering the coordination around the $Mn^{III}$ centers, thereby leaving the single-ion physics more-or-less unaltered. This allows for a detailed and systematic study of the way in which the individual-ion anisotropies project onto the molecular spin ground state in otherwise identical low- and high-spin molecules, thus providing unique insights into the key factors that control the quantum dynamics of SMMs, namely: (i) the height of the kinetic barrier to magnetization relaxation; and (ii) the transverse interactions that cause tunneling through this barrier. Numerical calculations are supported by an unprecedented experimental data set (17 different compounds), including very detailed spectroscopic information obtained from high-frequency electron paramagnetic resonance and low-temperature hysteresis measurements. Comparisons are made between the giant spin and multi-spin phenomenologies. The giant spin approach assumes the ground state spin, *S*, to be exact, enabling implementation of simple anisotropy projection techniques. This methodology provides a basic understanding of the concept of anisotropy dilution whereby the cluster anisotropy decreases as the total spin increases, resulting in a barrier that depends weakly on *S*. This partly explains why the record barrier for a SMM (86 K for $Mn_6$) has barely increased in the 15 years since the first studies of $Mn_{12}$-acetate, and why the tiny $Mn_3$ molecule can have a barrier approaching 60% of this record. Ultimately, the giant spin approach fails to capture all of the key physics, although it works remarkably well for the purely ferromagnetic cases. Nevertheless, diagonalization of the multi-spin Hamiltonian matrix is necessary in order to fully capture the interplay between exchange and local anisotropy, and the resultant spin-state mixing which ultimately gives rise to the tunneling matrix elements in the high symmetry SMMs (ferromagnetic $Mn_3$ and $Ni_4$). The simplicity (low-nuclearity, high-symmetry, weak disorder, etc..) of the molecules highlighted in this study proves to be of crucial importance. Not only that, these simple molecules may be considered among the best SMMs: $Mn_6$ possesses the record anisotropy barrier, and $Mn_3$ is the first SMM to exhibit quantum tunneling selection rules that reflect the intrinsic symmetry of the molecule.

# 1. Introduction

Now that the dust has settled on over 15 years of extensive study of the phenomenon of quantum tunneling of the magnetic moment (QTM [1]) of so-called single-molecule magnets (SMMs [2,3]), one arrives at two rather startling realizations: (1) the blocking temperature, $T_B$, below which a SMM can retain its classical up/down magnetic state, has barely increased from the value found for $Mn_{12}$-acetate back in 1993 ($T_B$ ~4K [4,5] for $Mn_{12}$ and ~4.5 K for $Mn_6$ [6]); and (2) the vast majority of studies of the quantum dynamics of SMMs have centered on the spin $S$ = 10 $Mn_{12}$ [3,4,7-9][†] and $Fe_8$ [10,11] molecules, in spite of considerable successes by chemists in synthesizing many other SMMs. The likely reasons for the latter realization will be outlined briefly below. Meanwhile, the reasons for first realization are also largely known, and will be discussed extensively in this article. In spite of this, strategies for substantially increasing $T_B$ have remained elusive.

Informed mainly by our own recent work on relatively simple, low-nuclearity molecule-based magnets [12-20], this perspectives article takes a broad view of our current understanding of magnetic bistability (in molecules) and QTM. Our primary aim is to highlight two important points: first, while adding spins to a molecule represents an obvious strategy for increasing $S$ and, thus, the barrier to magnetization relaxation,[‡] this leads also to a dilution/decrease of the molecular anisotropy, $D_{mol}$, which typically offsets much of the apparent gain due to increasing $S$; second, increasing nuclearity unavoidably results in increased complexity, e.g. many degrees of freedom (a large Hamiltonian matrix), many competing or frustrated interactions, increased disorder, etc. While complexity may be fascinating in other settings (e.g. in biology), it often obscures the inherently simple quantum physics underlying the dynamics of SMMs. For this reason, we have devoted much attention to low-nuclearity molecule-based magnets in recent years, e.g. $Mn_3$ ($S$ = 2 or 6 [19,21-23]), $Ni_4$ ($S$ = 4 [12-15,24]), $Mn_6$ ($S$ = 4 or 12 [18]). Ironically, one particular $Mn_6$ species now holds the record for the highest barrier to magnetization relaxation for a SMM ($U_{eff}$ = 86K [6]), while the barrier for the best $Mn_3$ SMM ($U_{eff}$ ~50K [19,25]) is nearly 70% of that of $Mn_{12}$ [4]. In other words, there is minimal gain upon increasing the nuclearity of molecules based on $Mn^{III}$ from three to twelve (at best a factor of two!). Meanwhile, studies of larger molecules have consistently yielded smaller barriers. This is due, in large part, to the magnetic anisotropy dilution that occurs upon increasing the number of metal centers in a cluster. This idea has been discussed recently by several groups [26-28], including the authors of this paper [12-20]. Here, we verify and expand upon these ideas, with particular emphasis on detailed comparisons that can be made with a wide body of available spectroscopic data.

So why have so many studies of magnetization relaxation in SMMs focused on just two molecules—$Mn_{12}$ and $Fe_8$? There are a multitude of reasons, some of which we briefly list here. First of all, the SMM and QTM phenomena were first discovered in $Mn_{12}$-acetate [29,30], while $Fe_8$ [10] was reported to display SMM properties [11] shortly after the groundbreaking QTM papers on $Mn_{12}$ appeared. Consequently, these two molecules have contributed to many of the most important discoveries associated with the magnetization dynamics of isolated SMMs, i.e. QTM, magnetic quantum phase interference (QPI) [31], magnetic avalanches [32,33], etc. Moreover, $Mn_{12}$ and $Fe_8$ are readily available (and easily synthesized) as good sized single-crystals with well defined shape/morphology, simplifying identification of the principal magnetic axes and, hence, alignment for magnetic studies. Crystals of both $Fe_8$ and $Mn_{12}$ are also stable in air and, although they exhibit some disorder, this is relatively weak and well characterized [34-38]. High blocking temperatures, particularly for $Mn_{12}$, further simplify studies within the blocked and pure tunneling regimes without the need for highly specialized low-temperature instrumentation. These factors have resulted in the research on $Mn_{12}$ and $Fe_8$ reaching a critical mass, particularly in the physics literature. Meanwhile, from the point-of-view of chemistry, $Mn_{12}$ is readily amenable to ligand substitution and crystallization from a variety of solvents, resulting in a very extensive family of $Mn_{12}$ SMMs [8], e.g. high and low symmetry, slow and fast relaxing, integer and odd-integer spin, etc.. This flexibility has, itself, yielded important insights into the magnetization dynamics associated with SMMs and, hence, to many more studies of $Mn_{12}$ compounds [8,9,39,40].

---

[†] Mostly $[Mn_{12}(O_2CCH_3)_{16}(H_2O)_4]\cdot 2CH_3CO_2H\cdot 4H_2O$, though more recent studies have focused on a range of closely related complexes with improved properties, see e.g. refs [8,9].

[‡] Note that the barrier to magnetization reversal for a SMM is given roughly by $S^2|D_{mol}|$.

In spite of the above, there are many aspects of the $Mn_{12}$ and $Fe_8$ SMMs that are far from ideal. Firstly, the magnetic interactions within these (and many other) ferri-magnetically coupled SMMs are inherently frustrated [4,41]. Second, the exchange and local spin-orbit anisotropies are not so different in strength, i.e. there is no truly dominant magnetic energy scale. These combined factors can give rise to a large density of low-lying spin states. Consequently, rigorous theoretical treatments should ideally consider all spin-spin interactions as well as the spin-orbit coupling at each metal center—thus, approximation carries substantial risk (see following section). Even when symmetry is taken into consideration, one finds that the idealized ($S_4$) $Mn_{12}$-acetate molecule requires four independent exchange coupling parameters and three differently oriented anisotropy tensors [4] (the same for $Fe_8$ [41]). This gives rise to an unmanageable number of unknown parameters that no realistic data set can completely constrain. Thus, it becomes a near impossible task to address important questions concerning the roles of, e.g., non-Heisenberg interactions or 4$^{th}$ (and higher) order local anisotropies. This is where the advantages of simple, low-nuclearity molecules such as high-symmetry ($C_3$) $Mn_3$ become very apparent, where one need consider only a single exchange constant, $J$, and a single anisotropy tensor, i.e. just three basic coupling parameters ($J$, $d_{ion}$ and $e_{ion}$), and a single matrix that specifies the orientation of the local anisotropy axes.$^§$ Note: hereon, we use lower case $d_{ion}$ and $e_{ion}$ to denote the single-ion 2$^{nd}$ order axial and rhombic zero-field splitting (ZFS) interactions, with upper case symbols reserved for the molecular ZFS parameters of the giant spin Hamiltonian.

There is another disadvantage that one invariably encounters in larger molecules—that is, disorder [14]. In particular, crystallization of larger molecules typically results in sizeable voids that are filled with loosely bound (volatile) solvent molecules. These solvents are often disordered and are easily lost via evaporation [42,43], leading to considerable strains in the molecular ZFS parameters [44]. Furthermore, in cases where intermolecular interactions are present, solvent disorder can give rise to additional complexity [19]. In contrast, a few smaller molecules (e.g. $Mn_3$, $Ni_4$) have been found to crystallize with no solvent [15,22,25]. Electron paramagnetic resonance (EPR) studies reveal dramatic differences in the spectra obtained from crystals containing no solvent, as compared to those that do: the former give extremely clean/sharp spectra in comparison to the latter; we note that such comparisons have been possible between molecules for which the magnetic cores are otherwise identical [14], i.e. they differ only in the identity of the ligand and whether or not the structure contains solvate molecules. These findings highlight the non-innocence of the solvents. Indeed, in many published examples, the distributions in spectral parameters caused by disorder are comparable to the parameters themselves. Such disorder undoubtedly influences the magnetization dynamics, particularly the quantum dynamics, which is so sensitive to minor symmetry-breaking perturbations. This has been most clearly demonstrated in $Mn_{12}$-acetate, where the disordered acetic acid solvent molecules have a profound influence on the QTM behavior [34-36]. Indeed, disorder is most likely the reason why QTM selection rules were only recently observed for the first time in a $Mn_3$ molecule containing no solvent [45,[46]]. Finally, not only is solvent disorder a problem in larger molecules, but the possibilities for disorder associated with the ligands, or even the magnetic core itself, also increase with increasing nuclearity [40]. Therefore, it is not surprising that spectroscopic studies to date on large molecules (nuclearity > 12) have so far yielded minimal information [47].

---

$^§$ Note that only a single matrix is required, because the three $Mn^{III}$ centers are related by the appropriate symmetry operation. Furthermore, the local axial direction can often be inferred from Xray data on the basis of the obvious Jahn-Teller elongation associated with the octahedral coordination around the $Mn^{III}$ ion.

## 2. Results and discussion
### 2.1 Giant-spin versus multi-spin phenomenology

For reasons outlined in the preceding section, analyses of most experiments on SMMs have typically been restricted to a giant spin description in which only the lowest-lying spin multiplet is considered ($S$ = 10 for both $Mn_{12}$ and $Fe_8$), yielding a set of phenomenological magnetic anisotropy parameters corresponding to the following zero-field spin-Hamiltonian [3,36,49]:

$$\hat{H} = D\hat{S}_z^2 + E\left(\hat{S}_x^2 - \hat{S}_y^2\right) + B_4^0 \hat{O}_4^0 + B_k^q \hat{O}_{4,6,8...}^q. \tag{1}$$

The first two terms represent the 2$^{nd}$ order axial and rhombic ZFS interactions, parameterized respectively by $D$ and $E$; the $\hat{S}_\alpha$ ($\alpha$ = $x$, $y$, $z$) are spin component operators. Subsequent terms correspond to 4$^{th}$ and higher order interactions: the $\hat{O}_k^q$ represent so-called Stevens operators of rank $k$, having $q$-fold symmetry ($q \leq k$) about the axial direction [48] . This approach assumes that $S$ is an exact quantum number and entirely ignores the internal degrees of freedom within the molecule, i.e. the relatively weak exchange couplings between the individual magnetic ions. Its advantage lies in the fact that it reduces the physics to a relatively small number of parameters which can often be determined accurately from spectroscopic measurements, e.g. inelastic neutron scattering (INS [38]) and EPR [36,37]. Indeed, most aspects of the QTM in $Mn_{12}$ and $Fe_8$ may be well understood considering the above Hamiltonian truncated to fourth order, i.e. just a g-tensor and three ZFS parameters for high-symmetry ($S_4$) $Mn_{12}$ [49,50] ($D$, $B_4^0$ and $B_4^4$, also $E$ if you take disorder in the acetate into consideration [36]), or five ZFS parameters for the lower symmetry (biaxial) $Fe_8$ [51].

The giant spin approach [Eqn. (1)] provides the added benefit of drastically reducing the dimension of the Hamiltonian matrix, thus enabling exact diagonalization on a standard PC. To put this into perspective, the matrix dimension associated with the twelve uncoupled Mn ions (eight $Mn^{III}$ with $s_A$ = 2, and four $Mn^{IV}$ with $s_B$ = 3/2) in $Mn_{12}$-acetate is $[(2s_A + 1)^8(2s_B + 1)^4]^2 = 10^8 \times 10^8$; for $Fe_8$, the matrix dimension is $10^6 \times 10^6$. Diagonalization of such matrices is a challenge even for the most powerful supercomputers. Even after reasonable approximation, one is left with a matrix of dimension $10^4 \times 10^4$ in the case of Mn$_{12}$ [4,52], not to mention two remaining exchange parameters and two ZFS tensors. By comparison, the giant spin Hamiltonian matrix for a spin $S$ = 10 object has dimensions of just 21 × 21, requiring only three ZFS parameters for $S_4$ symmetry (when restricting to 4$^{th}$ order). Within the giant spin approach, the advantages of studying simpler/smaller molecules are not quite so apparent, e.g. the matrix dimension for Mn$_3^{III}$ is 13 × 13, and a rigorous parameterization of the spectrum still requires at least three ZFS parameters, even for the highest symmetry ($C_3$) case [45]. It is when one treats low-nuclearity molecules within a multi-spin description that the advantages become significant, as we shall show below.

The giant spin approach has been hugely successful, accounting for many of the initial discoveries involving Mn$_{12}$ and Fe$_8$ with remarkable accuracy [3]. However, it fails to account for the internal degrees of freedom within a molecule in cases where the total spin ground state may lack rigidity. This was first illustrated for the case of the coupled relaxation associated with two weakly coupled Mn$_4$ SMMs (a dimer) [53]. Very recent studies have highlighted cases where the giant spin Hamiltonian [Eqn. (1)] fails completely to account for QTM resonances associated with individual molecules where, in addition to the usual change in spin projection, the total spin also fluctuates [54-58]. The reason for this failure is obvious: using Mn$_{12}$ as an example, the giant spin approach completely ignores ($10^8$ − 21) higher-lying spin states that can admix with the $S$ = 10 ground state. In spite of this, and as stated above, the 'effective' giant spin phenomenology continues to work very well for Mn$_{12}$ and Fe$_8$. However, one obtains a renormalized ZFS parameter set via such an approach [different from what would be expected if $S$ were exact—see section 2.3(c)], including, e.g., unphysical 4$^{th}$ and higher order terms [13-15,45,59]. Hence, it becomes a real challenge to correlate the results of EPR and QTM measurements to underlying details of the molecular structure, something that represents the ultimate goal of spectroscopists and synthetic chemists alike.

For the aforementioned reasons, it becomes abundantly clear that one can gain far deeper insight into the connection between physical properties (e.g. hysteresis loop step heights) and underlying molecular structure by considering a multi-spin Hamiltonian of the form [13,45]:

$$\hat{H} = \sum_i \sum_{j \neq i} J_{ij} \hat{s}_i \cdot \hat{s}_j + \sum_i \left[ d_i \hat{s}_{zi}^2 + e_i \left( \hat{s}_{xi}^2 - \hat{s}_{yi}^2 \right) \right]. \qquad (2)$$

The double summation parameterizes the isotropic exchange coupling, of strength $J_{ij}$,** between spins $i$ and $j$ in the molecule; $\hat{s}_i$ and $\hat{s}_j$ represent individual spin operators. The 2$^{nd}$ summation parameterizes 2$^{nd}$ order ZFS interactions associated with the individual ions, and may be compared directly to Eqn. (1). Application of Eqn. (2) is where one runs headlong into issues of matrix dimension and over-parameterization. Most importantly, this is where one clearly recognizes the huge advantage of studying simpler, low-nuclearity SMMs. For example, the matrix dimension for Mn$_3$ is just $5^3 \times 5^3 = 125^2$, in comparison to the $10^8 \times 10^8$ for Mn$_{12}$. Furthermore, only three parameters ($J$, $d_{ion}$ and $e_{ion}$) are required to completely specify the ZFS interactions, together with a single matrix that dictates the relative orientations of these interactions [45]. For $Ni_4$, which can crystallize with exactly the same $S_4$ symmetry as many $Mn_{12}$'s, the matrix dimension is just $(3^4)^2 = 81 \times 81$. Here again, only a handful of parameters (3, or 4 at most) and a single Euler matrix are needed to fully specify the ZFS interactions [13].

---

** We use the $J\hat{S}_1 \cdot \hat{S}_2$ convention rather than the -2$J\hat{S}_1 \cdot \hat{S}_2$ convention found in many chemistry journals. This should be taken into consideration when making comparisons between $J$ values quoted in this and other papers.

In the following sections, we make several mappings between the giant-spin [Eqn. (1)] and multi-spin [Eqn. (2)] phenomenologies, with emphasis on parameters that are important in the context of the physics of SMMs, i.e. the axial anisotropy ($D_{mol}$) which gives rise to the magnetic bistability, and the transverse interactions that cause QTM (e.g. $E_{mol}$). In this way, we can for example relate $D_{mol}$ (or $E_{mol}$) to $J$ and $d_{ion}$ (as well as $e_{ion}$). At every step, our findings are informed by real molecules that have been extensively characterized by magnetic and spectroscopic (EPR) techniques which are published elsewhere [12-19,22,25]. We focus primarily on the $Mn_3^{III}$ and $Mn_6^{III}$ SMMs, with occasional reference to $Ni_4^{II}$. In the following section, we begin by considering approximate methods based on the projection operator formalism [60]. This approach, which assumes $S$ to be exact, enables a very clear illustration of the anisotropy dilution alluded to in the introduction. In Section 2.3, we resort to matrix diagonalization in order to illustrate the consequences of spin state mixing. This is followed by Section 2.4, which discusses the overall implications in terms of QTM.

## 2.2 The connection between $D_{mol}$ and $d_{ion}$

If the isotropic exchange in Eqn. (2) is very strong in comparison to all other interactions, i.e. $J \gg d_{ion}$, then $S$ is an exact quantum number. In such a limit, one may project (add) the 2nd order anisotropy tensors of the individual magnetic ions onto any given spin manifold,[††] following techniques described by Bencini and Gatteschi [60]. The procedure begins by coupling two ions, then re-computing the anisotropy for the dimer in terms of the individual anisotropies of the constituent ions. One can then extend this technique, in principle, to any arbitrary number of particles by adding spins one at a time to obtain a trimer, tetramer, etc., re-computing the total anisotropy each time. This procedure turns out to be relatively straightforward for the case of ferromagnetically coupled molecules; the ferri- and antiferromagnetic cases are less trivial, because the result is dependent on the employed coupling scheme.

The ideas discussed above are laid out in more detail in refs. [16,17,61], which treat the cases of oxo-centered $Mn_3^{III}$ triangles and coupled hexanuclear $[Mn_3^{III}]_2$ in some detail (see structures in Fig. 1). We chose to focus on these two systems for a number of reasons: (1) one of the $Mn_6^{III}$ complexes holds the record for the highest magnetization relaxation barrier for a SMM [6]; (2) the $Mn_6^{III}$ complexes are remarkably similar to the oxo-centered $Mn_3^{III}$'s in that they are essentially comprised of two ferromagnetically coupled $Mn_3^{III}$ units (see Fig. 1), i.e. they may be viewed as a ferromagnetic $[Mn_3^{III}]_2$ dimer; (3) extensive families of both molecules exist (20-30 members in each case), differing only in the peripheral ligands and solvent molecules, resulting in relatively minor, albeit controllable perturbations to the structures [18-20,22,25]; and, most importantly, (4) one can vary the magnitude and sign of the exchange within the oxo-centered $Mn_3^{III}$ triangles [19], enabling studies of both low- and high-spin molecules ($S$ = 2 and 6 in the case of $Mn_3^{III}$, $S$ = 4 and 12 for $Mn_6^{III}$). Before summarizing the main findings of these studies, we point out several approximations and simplifying assumptions that were made in refs. [16-19]. Our main goal here was not to achieve precise agreement with experiment, but rather to provide a general understanding of the factors that contribute to the molecular axial anisotropy, $D_{mol}$, in polynuclear clusters, using $Mn_3^{III}$ and $Mn_6^{III}$ as examples. For this section (2.2) and the following one (2.3), our simplifying assumptions were as follows:

(1) We neglected dipolar contributions to the anisotropy, since these are typically an order of magnitude weaker than the $Mn^{III}$ single-ion anisotropies [62].
(2) We considered only axial single-ion ZFS of the form $d_i \hat{s}_{zi}^2$, where $s_i = 2$ and the $\hat{s}_i$ represent local spin operators associated with the individual $Mn^{III}$ centers.
(3) We assumed identical values for $d_i$ for all molecules ($Mn_3^{III}$ and $Mn_6^{III}$, high-spin and low-spin), i.e. we assumed that any structural modifications, even those that switch the sign of the exchange couplings ($J_{ij}$), do not significantly affect the spin-orbit coupling associated with the individual $Mn^{III}$ centers.
(4) We assigned identical $d_i$ values to each site in a given molecule.
(5) We assumed the axial directions (Jahn-Teller axes) to be parallel.
(6) We employed only a single value for $J$ on all bonds, switching only the sign on the relevant bonds for the low-spin cases (see below and ref. [16,17]).

---

[††] Note that, for clusters with more than two spins, additional quantum numbers are required to specify the symmetries of the different intermediate spin states ($0 \leq S < NS_i$, where $N$ denotes nuclearity and $S_i$ the spin associated with the single ions) obtained using different coupling schemes.

These simplifications conveniently reduce the problem to a single anisotropy parameter, $d_{ion}$, representing the axial ZFS associated with a single $Mn^{III}$ ion, and a single $J$ coupling constant. (3) and (4) are undoubtedly oversimplifications. However, our goal here was to obtain a simple correspondence between $D_{mol}$ and $d_{ion}$ (and, eventually, $J$ — see Section 2.3) which would not be possible otherwise. In this section, we shall see that this enables us to compare ($d_{ion}$ independent) ratios of $D_{mol}$ values for different complexes. (5) is a reasonable assumption based on Xray data [18,19,25], i.e. the Jahn-Teller axes are mostly reasonably parallel (in many cases within 20° of the average axial direction). Likewise, (2) is reasonable in the light of (5), if we are concerned only with $D_{mol}$. Finally, (6) is mostly irrelevant for the calculations considered here, which assume that $S$ is exact (i.e. the $J$'s are infinite) — only the sign of $J$ is important. We note that (6) will become more relevant when we consider weak $J$ coupling (Section 2.3).

The results of our calculations are summarized in Table 1. The most important findings are summarized in the final column which gives ratios for the molecular anisotropies, $D_S$, and magnetization barriers, $U_S$. As can be seen, upon increasing $S$ by a factor of 3 (by switching the sign of the exchange within the oxo-centered triangles), $D_{mol}$ decreases dramatically by a factor of somewhere between 5 and 6. Something similar happens upon increasing the nuclearity from three to six, although not quite as dramatic, i.e. $D_{mol}$ decreases by a factor of ~2.1. This 'dilution' of the anisotropy significantly offsets the effect of increasing $S^2$, i.e. one anticipates only a moderate increase in the barrier (a factor of ~1.45 – 1.75) upon switching from low- to high-spin, and a factor of less than two upon doubling the nuclearity (spin). The concept of $D$ decreasing as $S$ increases is not new [16-20,26-28]. However, our results paint a slightly more optimistic view in comparison to two previous works suggesting that $U$ does not increase upon increasing $S$ [26,27]. More importantly, we shall now see that the trends highlighted in Table 1 are borne out remarkably well by experiment. As summarized in Table 2, we compare our findings not only with magnetic studies, but also to an extensive body of spectroscopic (EPR) data.

Before discussing experimental trends, it is important to make several cautionary remarks. In general, $S = 2$ is not the anticipated ground state for antiferromagnetic $Mn_3^{III}$'s that possess an approximately equilateral triangular geometry (~same $J$'s on all three bonds, see Fig. 1). In fact, one expects a singlet, $S = 0$, ground state in this situation due to the frustration inherent in such a topology. As we shall see, reliable experimental data have been obtained only for those molecules in which the frustration has been mostly relieved, particularly in the case of EPR [19,25]. For the $Mn_6$'s, it is the ferromagnetic coupling between the halves of the molecule which naturally relieves the frustration within the triangular units (see Fig. 1). We have verified this by means of numerical diagonalization of Eqn. (2) (see Section 2.3) for which a well isolated $S = 4$ ground state is obtained in the strong $J$ coupling limit. Experiments support this finding, i.e. there are plenty of $Mn_6$ molecules for which data exist that yield good fits to $S = 4$ models [Eqn. (1)], including EPR spectra [18]. In contrast, an exhaustive study found only two examples of low-spin $Mn_3$'s (out of more than six studied) that yielded any useful information from EPR [19,63,64]. These turn out to be the more distorted molecules for which some, or all of the frustration is apparently removed. In spite of this, it has not been possible to fit the EPR spectra to Eqn. (1) in either instance. All that one can conclude is that the ground state is magnetic ($S \sim 2$) and that there exists appreciable easy-axis magneto-anisotropy. This is most likely due to the fact that the exchange is not strong (comparable to $d_{ion}$) in these molecules, resulting in considerable spin-state mixing, particularly in the antiferromagnetic cases.‡‡ In fact, the appreciable Ising-like character associated with the individual $Mn^{III}$ centers will tend to favor the $m_S = \pm 2$ ground state configuration [25]. Nevertheless, $S$ is never well defined for the low-spin $Mn_3$ ground states.

Table 2 includes 17 compounds out of over 50 published examples that are available to us. This choice of 16 was guided first and foremost by those molecules for which EPR data have been collected (10 out of the 17). Seven additional examples were judiciously chosen, paying special attention to the issues regarding spin frustration discussed above, i.e. we considered parameters such as $J$ and the proximity of low lying states, factors that influence the reliability of fits to magnetic data. Many compounds were inappropriate for this study, due either to unavailable or inconclusive data, e.g. intermediate or uncertain spin values. By the same token, plenty of compounds were left out of Table 2, even though their inclusion would only strengthen the conclusions of this study.

There are several comparisons that can be made, as shown in Table 1. The most robust are those that avoid consideration of the low-spin $Mn_3$ complexes. Therefore, we consider first $Mn_6$. The $D_{12}$ values (–0.43 cm$^{-1}$) obtained from magnetic studies are highly reproducible, even when considering all of the examples tabulated in ref. [18] (variation < 20%). The $D_{12}$ values obtained from EPR are slightly smaller, but they are essentially the same for the two compounds studied. It is important to note that the Hamiltonian used to fit the spectroscopic data includes a 4$^{th}$ order $B_4^0$ term [see Section 2.3(c)]. For this reason, one should not make direct comparisons between $D_{mag}$ and $D_{EPR}$. The main conclusion that one can draw from Table 2 is that the anisotropy does not vary significantly between the different $S = 12$ Mn$_6$ complexes. This statement is based on measurements of $D_{mol}$, not $U_{eff}$. As we shall see below, $U_{eff}$ does not provide quite such a reliable measure of the cluster anisotropy. Meanwhile, the $D_4$ values deduced on the basis of magnetic measurements exhibit some variation (from –1.2 to –1.6 cm$^{-1}$), while only a single value is available from EPR (although the data and analysis are robust [18]). We believe that the variation is more a reflection of experimental uncertainty, than any intrinsic behavior [see Section 2.3(b)]. Overall, one may conclude that the ratio $D_4/D_{12}$ is somewhere in the range from 3 to 4 (~3.6 from EPR). This is a little lower than the 5.4 – 6.4 range listed in Table 1.

---

‡‡ The degree to which projection methods are reliable is related ultimately to the symmetries and proximity of nearby spin states in relation to the ZFS within these states, hence the ratio $|J/d_{ion}|$ gives a rough measure of validity. Indeed, these methods are, in principle, exact in the limit $|J| \to \infty$. However, the symmetries of the unperturbed states also play a crucial (and less predictable) role. This explains why giant spin approximations work better for ferromagnetic examples and less well for molecules with antiferromagnetic couplings, because the density of low-energy spin states is considerably higher in the latter, particularly when viewed as 2D plots of energy versus spin projection (see e.g. Figs. 5 and 6 in ref. [25]), i.e. quantum fluctuations of the spin are much stronger in antiferromagnetic molecules.

Contrary to the small variation in $D_{12}$, the $U_{12}$ values are all over the map. However, this is well understood: under-barrier tunneling is quite strong in some Mn$_6$ complexes (due to low-lying $S < 12$ states—see [43,56]), leading to an apparent reduction in $U_{eff}$. These factors were not taken into consideration in the preceding calculations. We believe the unperturbed height of the barrier (~$DS^2$ + any higher order contribution) for a typical high-spin Mn$_6$ SMM to be somewhere in the 70 to 80 K range, in comparison to ~30 K for the $S = 4$ complexes. This implies a ratio $U_{12}/U_4 \sim 2.3 – 2.7$, which is a little larger than predicted in Table 1, as expected given that the $D_4/D_{12}$ ratio is smaller than expected, i.e. these two findings are consistent.

An obvious conclusion one could draw from the Mn$_6$ data is that the discrepancies between theory and experiment are due to the simplifying assumptions, particularly those concerning the single-ion anisotropy [65]. It is certainly possible, even likely, that $d_{ion}$ is influenced by the structural modifications that result in the switching from high- to low-spin. However, we believe that there is another crucial factor at play here: namely, the validity of the effective giant spin approximation [13]. As already stated, $S$ is only exact in the limit that $J >> d_{ion}$. Using data from Table 2 and the results from Table 1, one may estimate that $d_{ion}$ is of the order –2 to –3 cm$^{-1}$, i.e. $d_{ion}$ and $J$ have very similar magnitudes. Consequently, the giant spin approximation is likely to break down, resulting in a significant mixing of exited spin states into the assumed ground state multiplet. This will have the effect of renormalizing the data in Table 1. The only way around this is to use matrix diagonalization techniques, which we save until the following section where we demonstrate improved agreement with theory. The mixing is expected to be much more severe for the antiferromagnetic (low-spin) case; the ferromagnetic ground states (either $S = 6$ or $S = 12$) are rather well isolated by comparison, even in the many instances where $J \sim d_{ion}$ [56]. From this point-of-view, comparisons between high-spin Mn$_3$ and Mn$_6$ ought to give improved agreement. Indeed, they do! Based upon spectroscopic (EPR) data, Mn$_3$'s **6** and **7** have $D_6$ values of order –0.7 cm$^{-1}$, while their analogs **15** and **16** have $D_{12} \sim$ –0.35 cm$^{-1}$, i.e. $D_6/D_{12} \sim 2$ (note there are no Mn$_6$ analogs of **8**, **9** & **10**). This is in excellent agreement (to within 5%) with the value of 2.09 given in Table 1. The same holds for $U_{12}/U_6 \sim 80/45$, or 1.8(2). We thus believe that this lends strong support for the ideas laid out within this section, and validates the simple approach that we have adopted in order to illustrate the correspondence between $D_{mol}$ and $d_{ion}$ and the concept of anisotropy dilution.

We conclude this section by examining the available data for Mn$_3$. The only sensible comparison involves the $D_{mol}$ values deduced from magnetic studies. If one simply averages the available data, one obtains $D_2/D_6 \sim 4$. This is similar to the result for the Mn$_6$'s, both in terms of magnitude and in terms of the discrepancy from theory (Table 1). In other words, the Mn$_3$ data appear to further support our conclusions. However, it should be noted that there is a considerable experimental uncertainty associated with the measurements of $D_{mol}$. Meanwhile, only a single low-spin Mn$_3$ complex has so far been shown to exhibit slow relaxation at low temperatures [63]. The $U_{eff}$ value, which was inferred from hysteresis measurements, is rather small and may very well be indicative of under-barrier tunneling.

**2.3 The breakdown of the giant spin formalism**

**2.3(a) The effect on $D_{mol}$**

In the weak exchange limit, $J \sim d_{ion}$ (or $J < d_{ion}$), $S$ is no longer exact, i.e. the notion of well defined spin states ceases, and the projection methods described in the preceding section begin to lose validity. To circumvent this, one must use either perturbative or exact matrix diagonalization techniques; the former can be more intuitive, but the latter is simpler, provided the Hamiltonian matrix is not too large. As a first demonstration of the effect of $S$-mixing, we revisit the comparisons between high- and low-spin Mn$_6$. Fig. 2 plots $D_4/D_{12}$ as a function of the ratio $J/d_{ion}$, obtained via numerical diagonalization of the multi-spin Hamiltonian of Eqn. (2) using the sparse matrix algorithm [66] on a standard PC; for Mn$_6$, the matrix dimension is $(5^6)^2$ = 15625 × 15625. Precisely the same six simplifying assumptions concerning the single-ion ZFS and $J$ coupling were made, as outlined in the preceding section (2.2), and the $J$ coupling schemes are described in Refs. [16,17].

The first thing to note from Fig. 2 is that $D_4/D_{12}$ = 6.2 in the strong $J$ coupling limit. This falls slightly below one of the values obtained via the projection operator formalism, for which $D_4/D_{12}$ = 6.4 in the case of the antiferromagnetic coupling Scheme 1 (see Table 1). We suspect that this is because neither coupling scheme is exact, i.e. there is always some mixing between the different $S$ = 4 eigenstates corresponding to the two schemes. In the Mn$_3$ case, where there is no such ambiguity, we find exact agreement in terms of the ratio $D_2/D_6$ (and $U_6/U_2$) determined in the strong $J$ coupling limit on the basis of the matrix diagonalization and projection operator techniques [see Section 2.3(c)].[§§] Returning to Fig. 2, one sees that $D_4/D_{12}$ decreases significantly as $J$ decreases, displaying an approximately logarithmic dependence for $J/d_{ion}$ < 10, i.e. the ratio tends towards the experimentally determined range of 3 – 4. Very similar behavior is observed for Mn$_3$, where exact matrix diagonalization is trivial (matrix dimension is 125 × 125). We thus believe that Fig. 2 supports our previous assertion that the discrepancy between experiment and the simple calculations summarized in Table 1 can be attributed primarily to the breakdown of the giant spin formalism, i.e. $S$ is not exact.

In order to gain deeper insight into the reason for the decrease in $D_4/D_{12}$ seen in the main panel of Fig. 2, the inset separately plots the dependence of $D_4$ and $D_{12}$ on $J/d_{ion}$. As can clearly be seen, the variation is due entirely to $D_4$ (blue curve in upper inset). Before discussing this behavior in more detail, an important property of the ferromagnetic case merits further discussion. In the strong $J$ coupling limit, the fundamental excitations involve a uniform precession ($\Delta S$ = 0) of the total spin. For Mn$_6$, the lowest energy mode ($m_S$ = -12 to -11) has an energy $\Delta_0$ = $(2S – 1)D_{mol}$ = 23 × $D_{12}$ = $3d_{ion}$ in zero-field (Table 1). Remarkably, this is precisely the same as the lowest energy mode associated with an isolated Mn$^{III}$ ($s$ = 2) ion, i.e. $\Delta_0$ = $(2s – 1)d_{ion}$. Thus, we see that the fundamental excitation (with energy $\Delta_0$) associated with six (or, in fact, $N$) ferromagnetically coupled spins apparently does not depend on the strength of the coupling (provided one rigorously adheres to simplifying assumptions (1), (2) and (5) listed in Section 3). This is essentially a classical result which reflects the fact that the uniform (in-phase) precession of $N$ spins about a fixed axis is insensitive to the interactions between them, so long as the coupling is ferromagnetic and isotropic; this is analogous to Kohn's theorem for interacting electrons undergoing cyclotron motion [67]. On the basis of this result, the $J$ independence of $D_{12}$ (red line, inset of Fig. 2) makes sense because it was inferred directly from $\Delta_0/(2S-1)$. Not only does the value obtained in this way agree exactly with the result in Table 1 ($D_{12} = 0.1304d_{ion}$), it apparently holds for all values of $J$! As we shall see below, this contrasts the antiferromagnetic case, for which there is no simple classical description. It also does not hold for the higher order ferromagnetic modes.*** We will come back to the ferromagnetic case in Section 2.3(c), where we comment further on the procedure for deducing $D_{mol}$ directly from $\Delta_0$.

---

[§§] For the matrix diagonalization calculations, we force an $S \approx 2$ ground state by setting the coupling on one of the bonds to be ferromagnetic, thus relieving the frustration.

We now return to a discussion of the departure of $D_4$ from its strong coupling value (0.84$d_{ion}$, Scheme 1) for $J/d_{ion} < 10$. This is a manifestation of the fact that antiferromagnetic systems are intrinsically unstable against quantum fluctuations. Indeed, the antiferromagnetic eigenstates have no classical counterpart—they necessarily correspond to superpositions of products of the uncoupled basis states. Consequently, there is no correspondence between the coupled and uncoupled dynamics, as was found for the ferromagnetic case (note that, in the strong coupling limit, the uniaxial ferromagnetic ground state can be written as a pure product of the basis states, i.e. $|up\rangle|up\rangle|up\rangle$…). Thus, the low-energy modes associated with the antiferromagnetic molecules soften as $J$ softens, i.e. $\Delta_0$ and, therefore, $D_4$ (= $\Delta_0/7$) decreases as $J$ weakens relative to $d_{ion}$. We believe that it is this softening that is the primary reason for the discrepancy between the low- and high-spin data in Tables 1 and 2. This realization suggests that the $D_{mol}$ values associated with the low-spin molecules should be quite sensitive to $J$. We come back to this point in the following Section 2.3(b).

The situation discussed above may be further illustrated by comparing Figs. 5 and 6 in Ref. [25], which display 2D energy versus $m_S$ plots of spectra for $Mn_3$ complexes **4**, **8**-**10** obtained via exact diagonalization of Eqn. (2) (similar calculations for $Mn_6$ SMMs can be found in ref. [56]). The typical ratios of $J/d_{ion}$ used for these simulations are comparable, yet the nature of the low-energy states could not be more different. For the ferromagnetic examples, one clearly observes reasonably isolated parabolic bands (not only the ground state, but some excited ones as well) which may be associated with very definite spin values ($S = 6, 5, 4$…). Slight departures from parabolicity scale inversely with the proximity of excited $S = 5$ states [see Section 2.3(c)]. However, even for **8**, this departure is minimal. In contrast, it is impossible to determine any pattern amongst the low-energy states corresponding to the antiferromagnetic cases. Indeed, for the most frustrated example (not included in Table 2), one finds many closely spaced low-energy states having poorly defined $m_S$ values. The frustration is removed for **4** by setting the $J$ coupling on one of the bonds to ~0 (similar to the procedure outlined above [17]), resulting in the isolation of an $m_S = \pm 2$ ground state doublet. However, association of definite $S$ values with the low-energy states remains impossible, i.e. in spite of the removal of frustration, quantum fluctuations remain significant.

---

*** This is obvious when one recognizes that there are $S$ ($S - \frac{1}{2}$) non-degenerate zero-field modes associated with an integer (half-integer) molecular spin subjected to a cylindrically symmetric magnetoanisotropy, and only $s$ ($s - \frac{1}{2}$) modes for a single ion, where $S = Ns$ and $N > 1$.

**2.3(b) Experimental consequences of weak coupling and spin frustration**

By now, certain limitations of the experimental data sets in Table 2 have become apparent. The most obvious involves the reliability of the parameters obtained for the low-spin $Mn_3$ complexes (we discuss this further below). However, one also finds that this extends even to the ferromagnetic ($S = 6$) $Mn_3$'s. For example, spectroscopic (EPR) studies yield essentially identical $D_{EPR}$ values for complexes **6** and **7**, whereas the corresponding $D_{mag}$ fits vary by 60% (−0.77 and −0.48 cm$^{-1}$, respectively). These differences are due, in part, to a breakdown of the giant spin approximation, although intermolecular interactions are also strong for these molecules, representing another reason for the discrepancy. The $D_{mag}$ values for complexes **6** and **7** were obtained from fits of magnetization data to Eqn. (1) assuming a well isolated $S = 6$ ground state. In many of the cases considered here, this is not a good approximation, because there are low-lying ($S < 6$) states that also influence the field and temperature dependence of the magnetization, i.e. Eqn. (1) does not fully capture the physics, thus yielding parameters that do not correctly describe the ZFS within the lowest lying spin multiplet.

What about EPR spectroscopy? Low-lying states still affect the measurement. However, different excitations will, in principle, have different frequencies (or resonance fields) which are distinguishable by spectroscopic techniques such as EPR [15,68,69]. In fact, evidence for low-lying states is seen very clearly in $Ni_4$ (see Fig. 3(a) below), where a series of weaker resonances are apparent in between the stronger peaks. In many cases (including all of the $Mn_6$'s in Table 2 and the data in Fig. 3a), it is relatively straightforward to make assignments of the peaks that correspond to excitations within the lowest lying spin multiplet. These may then be fit to the giant spin model [Eqn. (1)] to yield reliable ZFS parameters. Indeed, as long as the assigned peaks may be fit to Eqn. (1) with a reasonable number of parameters (by convention, this usually means restricting to ~4$^{th}$ order), one can argue that this is a good approximation for EPR, even when it may not be the case when fitting magnetization data. If the peak positions cannot be fit to Eqn. (1), then one can conclude that the giant spin approximation has broken down completely (see e.g. [70]). We thus clearly see here the advantages of having access to high-frequency EPR data. Indeed, with the exception of the low-spin $Mn_3$ complexes (see below), all $D_{EPR}$ values given in Table 2 may be considered to be reliable, thereby lending even further support to the ideas discussed in this article. We emphasize that no comparable spectroscopic study presently exists in the literature.

The situation discussed above applies also to analysis of magnetic susceptibility data, where it is common practice to estimate exchange coupling parameters, $J_{ij}$, by means of fits to Eqn. (2) under the assumption that $d_{ion} = e_{ion} = 0$. This approach again runs into trouble when the $J_{ij}$ and $d_{ion}$ values are comparable, because the procedure ignores the ZFS within each spin multiplet. In other words, the problems associated with the determination of $D_{mag}$ and $J_{ij}$ from thermo-magnetic measurements are both related to separation of energy scales, i.e. the two methods described so far work only if spin multiplets are very well separated in comparison to the ZFS within the multiplets (i.e. $J \gg d_{ion}$). However, this is not the case for many of the complexes in Table 2. It is important to note also that EPR has been of limited use in terms of estimating $J_{ij}$ values, because selection rules normally forbid inter-spin multiplet transitions. It is, thus, more common to turn to INS for these purposes. However, we have shown recently that fits of EPR data to Eqn. (2) can give very reliable estimates of $J$ for molecules in which there is considerable $S$-mixing, i.e. $J \sim d_{ion}$ (see [13-15,22,25,45,71] for further details).

So how can one circumvent the problems discussed above in situations where there is no clear separation of energy scales? The only reliable approach is to fit all magnetic data (magnetization, susceptibility, low- and high-temperature) to the same multi-spin Hamiltonian [Eqn. (2)], including both the exchange (1$^{st}$ summation) and anisotropy (2$^{nd}$ summation) terms. This has been done for complexes **8** - **10**, yielding excellent agreement with EPR: the $d_{ion}$ values agree to within 6%, 4% and 2% for complexes **8**, **9** and **10**, respectively [22,25,45]. One can also deduce refined $D_{mag}$ values by mapping the spectrum obtained from fits to Eqn. (2) onto the giant spin Hamiltonian [Eqn. (1), see also following section]. These are the numbers given in Table 2 for complexes **8** – **10**. As can be seen, they agree with EPR to within ~6%. In other words, by using the correct physical model, one can achieve excellent agreement between fits to magnetic and spectroscopic data. It is much more difficult to follow this approach in the case of larger molecules, in part because of the computational difficulties associated with the diagonalization of large matrices, but primarily because the number of fit parameters inevitably increases. The information content associated with a typical set of magnetic measurements is basically the same regardless of the size of molecule. Thus, one can clearly see that the reliability of fits to Eqn (2) will suffer as the number of free parameters increases.

It is evident that the giant spin phenomenology improves for larger ferromagnetic SMMs. For example, fits to Eqn. (1) for Mn$_6$ complexes **15** - **17** yield identical $D_{mag}$ values and, although the $D_{EPR}$ values for **16** and **17** are smaller, they are the same for both. The differences between $D_{mag}$ and $D_{EPR}$ are due in part to the inclusion of 4$^{th}$ order anisotropy in the EPR analysis. However, EPR and INS [72] data do suggest lower overall anisotropy in comparison to the magnetic measurements, i.e. $U_{EPR}$ ~ 75 K. In the following section we discuss further reasons why magneto-thermal (and EPR) studies of ferromagnetic SMMs can be expected to give good agreement with the giant spin model. However, purely ferromagnetic polynuclear transition metal clusters are unfortunately rather rare, particularly larger ones.[†††]

The ideas discussed in this section are well known in the SMM community. However, we feel it worthwhile to discuss them here in order to give some context to the present study. We have clearly demonstrated limitations of the standard methodologies used to analyze magnetic data, even when considering apparently well behaved ferromagnetic Mn$_3$ SMMs, i.e. ones that display very clean EPR spectra and beautifully sharp hysteresis loop steps due to QTM, all of which can apparently be described very well according to the giant spin framework of Eqn. (1). The situation is far worse in the case of frustrated, or partially frustrated antiferromagnetic molecules. The low-energy spectra consist of many states with ill-defined $S$ and $m_S$ quantum numbers. EPR data cannot be fit to Eqn. (1). All that one can conclude is that these molecules possess significant magnetoanisotropy. We make no attempt here to analyze the situation further for the $S$ = 2 Mn$_3$ complexes, and it is for this reason that we devoted little discussion to these results in preceding sections. In the case of the low-spin ($S$ = 4) Mn$_6$ complexes, the frustration is partly relieved. Nevertheless, just one example so-far gave acceptable EPR spectra (out of four studied) that can be satisfactorily fit to Eqn. (1) with $S$ = 4.

---

[†††] This can be attributed to the fact that the superexchange between 3$d$ ions bridged by non-magnetic atoms tends to favor antiferromagnetic coupling.

## 2.3(c) Higher-order molecular anisotropy

For the remainder of the article, we focus on ferromagnetic molecules, with emphasis on SMM and QTM behavior. In order to simplify matters, we consider the simplest possible examples: ferromagnetic $Mn_3^{III}$ ($S$ = 6) [19,25], as well as some older results on $Ni_4^{II}$ ($S$ = 4) [24]. If one takes the view, as we did in Section 2.3(a), that $D_{mol}$ is the only relevant/finite ZFS parameter, then it may be directly related to $\Delta_0$, i.e. $D_S = \Delta_0/(2S - 1)$. However, calculations performed in the intermediate and weak coupling regimes reveal that a single $D_{mol}$ parameter cannot satisfactorily account for all of the energies of the (2$S$ + 1) low-lying states corresponding to the effective spin, $S$, ground state [13,45]. This is also backed up by many spectroscopic studies (EPR, INS and QTM) of real SMMs which demonstrate a clear need to include higher-order terms in any effective giant spin parameterization according to Eqn. (1) [3,13,45,49-51]. This is clearly illustrated by results obtained for $Ni_4$ (Fig. 3a) and $Mn_3$ (Fig. 3b) where a $B_0^4$ axial term must be included in the giant-spin Hamiltonian [Eqn. (1)] to account for the irregular spacing of the EPR peaks in $Ni_4$ and the temperature shift of the QTM resonances in $Mn_3$. The nature of these higher-order terms is such that they, too, contribute to $\Delta_0$, albeit that their contribution is typically small in those cases for which the giant spin approximation works well. Nevertheless, as soon as one begins to allow for high-order molecular ZFS, the $J$-independence of $D_{mol}$ (Fig. 2, lower inset) ceases, even though $\Delta_0$ remains $J$-independent.

The origin of 4$^{th}$ and higher-order molecular ZFS has been discussed at considerable length in refs. [13-15,45] and, more recently, in ref. [59]. We summarize the arguments here, focusing first on the axial terms (those containing only even powers of $\hat{S}_z^2$). We begin by making the assumption that the single-ion ZFS tensors contain only 2$^{nd}$ order terms. This approximation turns out to be exact for $S$ = 1 $Ni^{II}$ [12], because the single-ion matrix dimension is just 3 × 3, i.e. one expects just two 2$^{nd}$-order ZFS interactions of the form $d_{ion}\hat{s}_z^2$ and $e_{ion}(\hat{s}_x^2 - \hat{s}_y^2)$. For $Mn^{III}$, 4$^{th}$-order terms are allowed, but they are far too small to account for the observed molecular values [73]. Based on the matrix addition methods described in Section 2.2, one simply expects the single-ion anisotropies to project onto the exact spin ground state, i.e. one anticipates *only* 2$^{nd}$-order molecular ZFS, parameterized by $D_{mol}$ and $E_{mol}$. As noted above, such a parameterization fails to account for experimental observations (Fig. 3). For $Ni_4^{II}$, the *only* possible way out is to relax the strong $J$, exact $S$ assumption. We shall see that this is also the case for $Mn_3$ and, likely, most other SMMs. One can debate about whether this signifies a breakdown of the giant spin approximation. The fact is that the data displayed in Fig. 3 fit very well to Eqn. (1), so the approximation is fine. However, there is no immediately obvious correspondence between the molecular and single ion physics. It is, in fact, the finite exchange coupling that provides the additional freedom (parameter), i.e. $S$ is no longer rigid or exact. In this sense, one starts to see the limitations of the giant spin phenomenology, even if the approximation is OK. As $J$ decreases/softens relative to $d_{ion}$, the separation between ground and excited state multiplets decreases relative to the ZFS within each multiplet. These states also start to mix and interact, i.e. $J$ begins to influence the ZFS within each effective spin state. It is this interaction that gives rise to the higher-order molecular ZFS terms [13]. Eventually, if $J$ becomes sufficiently weak, excited states will overlap with the ground state. At this point, the giant spin approximation breaks down completely [54]. For example, the mixing with excited states can provide new relaxation pathways that result in an apparent reduction of the effective anisotropy barrier, $U_{eff}$

[43]. Indeed, this is precisely the reason why complex **15** ($J = -1.8$ cm$^{-1}$) has a significantly lower effective barrier in comparison to **16** and **17** ($J = -3.2$ cm$^{-1}$), even though the $D_{mol}$ parameters are apparently identical ($-0.43$ cm$^{-1}$) [56].

One can make a direct mapping between the molecular ZFS parameters ($D$, $B_4^0$, etc.) and the microscopic parameters ($J$, $d_{ion}$, etc., as well as the orientations of the ZFS tensors) by computing a given spectrum via diagonalization of Eqn. (2), then fitting the lowest lying states to Eqn. (1). Fig. 4 plots several such mappings corresponding to the effective $S = 6$ ground state for high-symmetry ($C_3$) Mn$_3^{III}$ (complexes **6-10**). The EPR barrier, $U_{EPR}$ (Fig. 4c), is defined as the energy difference between the ground state doublet ($m_S = \pm 6$) and the highest lying level ($m_S = 0$) associated with the $S = 6$ manifold. The calculations were again performed according to most of the simplifying assumptions listed in Section 3, with fixed $d_{ion} = -3.38$ cm$^{-1}$. In actual fact, the darker curves in each figure include a small tilting ($\alpha = 8.5°$) of the single-ion axial directions (local $z$-axes) away from the average molecular $z$-axis, i.e. we relaxed assumption #5. The tilting was added in such a way to preserve the trigonal symmetry of the molecule (see Ref. [45]); we discuss the consequences of this tilting in more detail at the end of this section and in the following section dealing with QTM.

The key result in Fig. 4 is the $J^{-1}$ dependence of $B_4^0$. In the infinite $J$ limit, the molecular ZFS is described entirely by $D_6 = 3d_{ion}/11$ (Table 1). Consequently, both $S$ and $m_S$ are exact (provided $e_{ion} = 0$), leading to the simple, well-known result, $E(m_S) = D_6 m_S^2$, i.e. an exactly parabolic $m_S$ dependence of the eigenvalues associated with the ground state manifold. As $J$ decreases, the mixing/interaction between spin states causes a deviation from parabolicity, as seen in the inset to Fig. 4(c); this of course assumes that one continues to label states according to exact $m_S$ values. The non-parabolicity is captured within the giant-spin picture [Eqn. (1)] via $B_4^0$ and higher order terms (see Fig. 5). From a perturbative point-of-view, one can easily see that the strength of the interaction between multiplets should scale inversely with their proximity [$E(S') - E(S)$], hence the $J^{-1}$ dependence $B_4^0$. The higher-order corrections appear in successive orders of perturbation, such that $B_6^0 \propto J^{-2}$, $B_8^0 \propto J^{-3}$, $B_{10}^0 \propto J^{-4}$ (see Fig. 5). In fact, if one re-scales the higher order interactions according to their contribution, e.g. to $\Delta_0$ (not shown in Fig. 5), one finds that it becomes necessary to consider all possible orders once the ratio of $J/d_{ion}$ reduces much below unity, i.e. the contribution of 6$^{th}$ and higher order terms to $\Delta_0$ become comparable for $J/d_{ion} \sim 0.5$. In this limit, one might say that the giant spin approximation has completely broken down.

As can be seen in Fig. 4, the full parameterization according to Eqn. (1) results in an obvious $J$-dependence of $D_6$, in contrast to the result presented in the lower inset to Fig. 2 (red line): the value of $D_6$ decreases by about 5% over the calculated range, as does $U_{EPR}$. The nature of the 4$^{th}$ and higher order Stevens operators is such that their effect on the overall barrier is weak[‡‡‡]—hence, the observed reduction in $U_{EPR}$ is due almost entirely to the variation in $D_6$. However, these same terms *do* contribute significantly to the relative spacing between $m_S$ levels within a given spin manifold, leading, e.g., to the uneven separation of the EPR peaks observed in Fig. 3(a). In fact, for parallel single-ion $z$-axes, the effect of $B_4^0$ (+ higher order terms) completely offsets the decrease in $D_6$ when it comes to the ground state splitting, i.e. $\Delta_o$ remains $J$-independent (not shown). This is not the case when the $z$-axes are tilted (see below).

---

[‡‡‡] Although, e.g., $\hat{O}_4^0$ (= $[35\hat{S}_z^4 - \{30S(S+1) - 25\}\hat{S}_z^2]$) does not contribute significantly to $U_{EPR}$, the non-parabolic $35\hat{S}_z^4$ component of the interaction does. Indeed, this purely quartic term contributes 20% of the barrier in the best know SMM, Mn$_{12}$-acetate.

We now see that weak $J$ coupling affects not only the antiferromagnetic molecules, but the ferromagnetic ones as well, i.e. $D_6$ and $U_{EPR}$ decrease by ~4% relative to the infinite $J$ (exact $S$) values for $J/d_{ion}$ = 1. However, the effect is much stronger for the antiferromagnetic molecules, i.e. a 25% reduction is observed in Fig. 2 for a comparable reduction in $J$. Consequently, as $J$ softens, one continues to expect a significant reduction of $D_2/D_6$ (and $D_4/D_{12}$) relative to the strong coupling calculations (Table 1), in agreement with experiment (Table 2). It is worth noting that magnetic measurements of $D_{mol}$ are usually conducted at low-temperatures in order to avoid population of excited $S$ states. This has the obvious effect of weighting the analysis towards the lowest energy portion of the spectrum. As we have seen, the splittings between these levels (e.g. $\Delta_0$) are relatively insensitive to $J$ in the case of ferromagnetic molecules. Thus, one may expect such analyses to approximate quite well to the strong coupling $D_{mol}$ values given in Table 1. These considerations reinforce the notion that the giant spin phenomenology works rather well for most ferromagnetic SMMs. However, it breaks down when excited states overlap significantly with the ground state, as is the case e.g. for complex **15** [56].

Very detailed spectroscopic studies (multi-frequency, multi-orientation single-crystal EPR) on the simple molecules $Mn_3^{III}$ ($S$ = 6) and $Ni_4^{II}$ ($S$ = 4) have provide remarkable support for the ideas outlined in this section. For example, in the $Ni_4^{II}$ case, a multi-spin analysis [Eqn. (2)] of the EPR data yields single-ion ZFS parameters and exchange constants that are in excellent agreement with independent measurements of the same quantities; for further details, see refs. [13-15]. Studies of $Mn_3$ complexes **6-10** [19,25] also beautifully confirm the predicted $J^{-1}$ dependence of $B_4^0$, as seen in Fig. 4(a), where the data are simply superimposed on the theoretical expectation: $B_4^0$ was determined from fits of EPR data to Eqn. (1); $J$ was determined independently from magnetic susceptibility measurements (Table 2); and $d_{ion}$ was determined from fits of the EPR data to Eqn. (2). It is important to note here that the agreement improves substantially when the single-ion easy ($z$-) axes are tilted by $\alpha$ = 8.5°. As we shall see in the following section, this tilting is also essential for accounting for QTM selection rules observed in low-temperature hysteresis measurements [45]. Furthermore, X-ray studies indicate that the Jahn-Teller axes associated with the $Mn^{III}$ ions are tilted by 8.5° relative to the crystallographic $z$-axis [25]. Indeed, $\alpha$ was set to this value for the calculations, as opposed to allowing adjustment in order to improve fits to experimental data. One may question using the theory that we are ultimately testing [Eqn. (2)] in order to obtain values for $d_{ion}$, i.e. there is an obvious correlation between $d_{ion}$ and $B_4^0$ (or $J$). However, this correlation is weak: varying $J$ produces at most a 3-4% renormalization of the $d_{ion}$ values obtained in this way (note that the error bars on $J/d_{ion}$ are ~10%). In other words, for a given value of $d_{ion}$, the appearance and variation of $B_4^0$ is due exclusively to $J$.

The high-symmetry ferromagnetic $Mn_3$ molecules provide a model system with which to test the ideas laid out in this section, requiring the addition of only a single additional parameter, $J$ (as opposed to multiple higher order $B_k^q$ coefficients). There is also very little variation in $d_{ion}$ among two distinct families of such complexes (**6** & **7** [19] and **8-10** [25]), which makes for a cleaner comparison. The fact that the experimental data agree so well with the calculations (as well as the results for $Ni_4^{II}$ [13-15]) surely rules out more exotic proposals [74]. Moreover, as we shall see in the following section, further support is provided by QTM studies [45], which allow for investigation of a mapping between the off-diagonal (tunneling) ZFS interactions obtained via the multi-spin and giant-spin methodologies. One other notable aspect of the ferromagnetic $Mn_3$ complexes considered in this section is the absence of lattice solvate molecules in all but one case (complex **6**) [19,25]. This results in some of the sharpest EPR spectra obtained for any SMM, enabling very accurate determinations of $B_4^0$ (or $J$). It is no coincidence that the one outlying data point in Fig. 4(a) corresponds to complex **6**, which gave broad EPR peaks with multiple fine structures, adding to the uncertainty in the determination of $B_4^0$. This is why the vertical error bar is quite significant for this one molecule. Here again, we see the advantage of studying simple SMMs with lattices that do not contain solvents [14]; the detailed analysis presented here would likely have been impossible otherwise.

We conclude this section by discussing the effect of tilting the single-ion easy ($z$-) axes on the ground state splitting $\Delta_0$, which serves as a natural segue to the following section dealing with QTM. The parallel $z$-axes case results in a $J$-independence of $\Delta_0$, which we have verified for several different ferromagnetic molecules (see e.g. Fig. 2). This was discussed quasi-classically above as being due to the insensitivity of $N$ spins precessing uniformly about parallel axes to the interactions between them. This no longer holds if the local $z$-axes are not parallel: a given spin will precess about its own $z$-axis in the weak coupling limit ($J \to 0$), whereas it will precess instead about the average (molecular) axial direction in the strong coupling limit. In other words, the local basis (specifically, the quantization axis) for a given spin rotates towards the molecular $z$-axis as the exchange coupling increases. Moreover, the projection of the individual ion anisotropies leads to a rescaling of $\Delta_0$, or to a $J$-dependence of the classical precession frequency. This is borne out by the exact diagonalization calculations (not shown). As seen in Fig. 4, the tilting also results in a very obvious decrease in $D_6$ and an increase in $B_4^0$; in effect, the tilting results in a further departure from the strong coupling, exact $S$ situation, i.e. it leads to increased $S$-mixing and a further break down of the giant spin phenomenology.

### 2.4 Implications for QTM

The $Ni_4$ and $Mn_3$ SMMs discussed in the preceding section share an important fundamental characteristic—their high symmetries forbid 2nd order transverse (tunneling) molecular ZFS interactions. Thus, one immediately sees that tunneling in these systems cannot result from a simple transverse projection of the 2nd order single-ion anisotropies ($d_{ion}$ and $e_{ion}$) onto the molecular spin ground state. Yet both complexes exhibit very clear QTM behavior. Indeed, the pure QTM rate is extremely fast in the case of $Ni_4$ (~10 MHz in zero-field). The explanation again lies in the emergence of higher-order transverse anisotropy as a result of mixing between spin multiplets. For the case of $Ni_4$ ($S_4$ molecular point group symmetry), we refer the reader to a series of articles [12-15] where this has been discussed in great detail. However, the more recent example of $Mn_3$ ($C_3$ point group) has provided additional insights into the origin of transverse anisotropy and its influence on QTM [45]. We therefore focus on this example here.

Low-temperature magnetization measurements on complexes **8** and **9** [45] yield hysteresis loops with extremely sharp QTM steps, indicative of the exceptionally high quality of these compounds which do not contain any lattice solvent molecules. For fields parallel to the easy-axis and temperatures above ~1.3 K, more-or-less evenly spaced QTM steps are observed (period ~0.8 T), starting from zero field. However, upon reducing the temperature, the $k = 1$

QTM resonance vanishes completely, indicating unmeasurably weak tunneling between the lowest metastable state ($m_S$ = +6), and the $m_S$ = –5 level of the stable well [see Fig. 3(b)], even though QTM is observed at the resonances immediately on either side ($k$ = 0 and 2) down to the lowest temperatures studied. This suggests that the intrinsic[§§§] tunneling matrix element associated with this level crossing vanishes. Indeed, on the basis of molecular symmetry, one does not expect tunneling at this particular level crossing. One can understand this within the giant spin formalism as being due to the fact that the ZFS interactions appropriate to the $C_3$ point group should only involve Stevens operators with even $k$ and $q = 3n$ ($n$ = integer), i.e. $\hat{O}_4^3$, $\hat{O}_6^3$, $\hat{O}_6^6$, etc. The spin raising and lowering operators ($\hat{S}_+$ & $\hat{S}_-$) occur only in powers of $q$ in these interactions. Consequently, they connect only states that differ in spin projection, $\Delta m_S$, by $q = 3n$. Clearly, the $m_S$ = +6 to –5 crossing ($|\Delta m_S|$ = 11) is not connected by these interactions. The same symmetry arguments apply even if the giant spin approximation breaks down. Thus, intrinsic ground state tunneling at the $k$ = 1 resonance is strictly forbidden. On the other hand, thermally activated tunneling is allowed at the $m_S$ = +5 to –4 level crossing ($|\Delta m_S|$ = 9) immediately above the $m_S$ = +6 to –5 crossing. It is for this reason that a QTM step appears at elevated temperatures. Several level crossings have been highlighted in Fig. 3(b) according to whether tunneling is forbidden (red circles) or allowed (blue circles) for the $C_3$ symmetry.

The vanishing of QTM at the $k$ = 1 resonance in complex **8** as T → 0, together with the simultaneous observation of the $k$ = 0 resonance (note, $|\Delta m_S|$ = 12 for this transition), represents the first clear-cut manifestation of an SMM obeying a quantum mechanical selection rule expected on the basis of its molecular symmetry. However, the above considerations indicate that one should not expect relaxation at the $k$ = 2 resonance either as T → 0 [see Fig. 3(b)], yet a step is always seen. This apparent conflict provides an outstanding opportunity to gain deeper insights into the reasons why intrinsically forbidden QTM resonances are nonetheless observed. To do so, we revert to use of Eqn. (2), using the following parameters: $d_{ion}$ = –2.91 cm$^{-1}$, $e_{ion}$ = 0.62 cm$^{-1}$, and isotropic $g$ = 2.00 for the $s$ = 2 Mn$^{III}$ ions, together with a single isotropic coupling parameter $J$ = –3.38 cm$^{-1}$. $d_{ion}$, $g$ and $J$ were chosen so as to accurately account for the positions of the QTM steps, while $e_{ion}$ was set to a value consistent with the measured relaxation rates. Finally, we ensured that the transverse single-ion ZFS interactions were oriented in a manner consistent with the trigonal symmetry (see [45] for details).

The key findings obtained on the basis of the above parameterization are summarized in Fig. 6, which plots the transverse applied field dependence of several tunnel splittings involving the metastable $m_S$ = +6 state, i.e. $k$ = 0, 1, 2, and 3. We focus first on the thin curves which were simulated assuming that the single-ion $z$-axes are parallel (assumption #5 from Section 2.2—see also Fig. 4). With the exception of the $k$ = 0 simulation, all of the tunnel splittings vanish at zero applied transverse field ($B_T$ = 0). This immediately suggests the absence of ZFS interactions connecting $m_S$ = +6 to –3 ($|\Delta m_S|$ = 9) levels. On the other hand, the $k$ = 0 splitting ($|\Delta m_S|$ = 12) is significant and essentially field-independent over the range of transverse fields shown in the figure. These observations suggest that only Stevens operator terms with $q = 6n$ emerge (e.g. $B_6^6 \hat{O}_6^6$) in the giant spin parameterization. Indeed, a mapping between the two models [Eqns. (1) and (2)] rigorously confirms this assertion [75]. Therefore, this situation cannot account for the T → 0 tunneling either at $k$ = 2 or 3! This can be understood as being a consequence of the 2$^{nd}$ order nature of the single-ion ZFS interactions such that the second summation in Eqn. (2) cannot distinguish between ±$x$ (likewise for ±$y$). This, together with the parallel arrangement of the single-ion tensors, results in the molecular Hamiltonian having $C_{3v}$ symmetry, rather than $C_3$, i.e. the addition of three mirror planes. Hence, a six-fold symmetry emerges ($q$ = 6$n$) as opposed to three-fold.

---

[§§§] Extrinsic tunneling interactions due to disorder and random electron/nuclear dipolar fields will never strictly go to zero (see below).

As noted in the previous section, overall agreement between experiment and simulation improves if one includes a tilting of the single-ion easy-axes. It was also noted that additional state mixing occurs as a result of this tilting. Precisely the same situation is found here, as illustrated by the thick curves in Fig. 6, which were generated including a tilt angle of $\alpha$ = 8.5° (consistent with the $C_3$ symmetry and Xray data). The $k$ = 0 resonance is barely affected by this tilting, because the relevant levels are already split by the $B_6^6 \hat{O}_6^6$ interaction. The $k$ = 1 and 2 splittings are affected ($k$ = 2 more so), but they continue to vanish as $B_T$ → 0, as they should do on the basis of the molecular symmetry. On the other hand, the $k$ = 3 resonance now has a huge associated tunnel splitting, i.e. the three-fold ($q$ = 3$n$) symmetry emerges, allowing Stevens operator terms such as $B_4^3 \hat{O}_4^3$ and $B_6^3 \hat{O}_6^3$. Tilting the single-ion axes breaks the mirror symmetry associated with the $C_{3v}$ point group, resulting in the Hamiltonian attaining the true chiral $C_3$ symmetry of these molecules. Once again, a mapping between the two models [Eqns. (1) and (2)] rigorously confirms this assertion [75].

We are now in a position to understand the experimental observations, and to draw conclusions about QTM selection rules in other SMMs. It is notable that the inclusion of tilting in Eqn. (2) resulted in an appreciable modification of the transverse field dependence of the tunnel splittings associated with the nominally forbidden $k$ = 1 and 2 resonances. However, the effect is far more pronounced for $k$ = 2 for small transverse fields. Indeed, $\Delta_{k=2}$ approaches the level of $\Delta_{k=0}$ for values of $B_T$ of order 25 mT. Meanwhile, $\Delta_{k=1}$ does not change so much over the comparable field range. This is significant, because electron dipolar field fluctuations of this order can be inferred from the widths of the hysteresis loop steps. Therefore, these dipolar fluctuations, which are ever present in real crystals, can apparently stimulate tunneling at the $k$ =2 resonance, while they are harmless in terms of tunneling at $k$ = 1. Another possible source of weak transverse fields is disorder, which can give rise to weak random distributions in the orientations of the molecular ZFS tensors so that application of a longitudinal field results in small local transverse components as well. We do not believe this to be a factor for the Mn$_3$ complexes considered here. Nevertheless, this likely represents an additional source of symmetry breaking in many other SMMs [37].

A final point worth noting in this section involves experimental comparisons between Eqns. (1) and (2). So far, it has not been possible to detect the transverse ZFS interactions for the Mn$_3$ complexes via EPR, whereas the $B_4^4 \hat{O}_4^4$ interaction is very apparent in Ni$_4$ [13,15]. In fact, it is possible to achieve excellent fits of high-field EPR data to both Eqn. (1) and (2) for Ni$_4$. However, comparisons of tunnel splittings obtained via the two models using the parameters obtained from EPR yield significant differences [15]. This suggests a crucial limitation (indeed, a breakdown) of the giant spin approximation [13]. In other words, a particular parameterization may perfectly account for the high energy physics (i.e. EPR transitions observed in the 10$^{11}$ Hz range), yet the same parameterization may completely fail to properly predict the low-energy physics (i.e. QTM splittings of order < $10^6$ Hz). The reason is because one typically truncates the analysis of EPR data to $4^{th}$ order, yet higher order terms may play a crucial role in the very delicate quantum dynamics.

## 3. Conclusions

We have presented a detailed comparisons between the giant-spin [Eqn. (1)] and multi-spin [Eqn. (2)] phenomenologies, with emphasis on parameters that are important in the context of QTM in SMMs, i.e. the axial anisotropy ($D_{mol}$) which gives rise to the magnetic bistability, and the transverse interactions that cause QTM (e.g. $E_{mol}$). In this way, we have for example been able to relate $D_{mol}$ (or $E_{mol}$) to the exchange coupling $J$ between constituent ions, and the local anisotropies associated with these ions, i.e. $d_{ion}$ (as well as $e_{ion}$). At every step, our findings have been informed by simple molecules that have been extensively characterized by magnetic and spectroscopic (EPR) techniques. The giant spin methodology provides a simple understanding of the concept of anisotropy dilution whereby the cluster anisotropy decreases as the total spin increases. This results in a barrier, $U$, that does not increase in proportion to $S^2$, as one might naively expect. Nevertheless, the experimental data and numerical calculations presented in this article suggest that one can achieve a moderate increase ($U \sim S^1$) by maximizing the spin associated with a particular cluster. This contrasts several recent studies claiming that $U$ should not increase with increasing $S$ [26,27]. Overall we find remarkably good agreement between theory and experiment, even for the simplest of calculations [Eqn. (1)]. Refinements of these methods, which take into consideration the internal degrees of freedom within a SMM [Eqn. (2)], lead to improved agreement with experiment and to deeper insights into the interplay between exchange and local anisotropy, and the resultant state mixing which ultimately gives rise to QTM in high symmetry SMMs (ferromagnetic $Mn_3$ and $Ni_4$). It should be emphasized that the availability of very detailed high-frequency EPR and low-temperature hysteresis data provided particularly stringent tests of the theoretical framework laid out in this article.

An overriding theme concerns the immense value of studies of relatively simple molecules such as the ones highlighted in this work. Indeed, one has some chance of maintaining a level of control over key structural parameters in simpler clusters [18,19], leading e.g. to: high symmetries ($Ni_4$ & $Mn_3$); minimal disorder; maximization of the molecular spin, $S$; and maximal projection of the available anisotropy. These factors likely explain why efforts aimed at producing larger SMMs have so-far failed to improve upon the original $Mn_{12}$-acetate. In contrast, $Mn_6$ now possesses the record anisotropy barrier for a SMM [6], while $Mn_3$ has a barrier that is 2/3 that of $Mn_{12}$-acetate and is the only SMM to exhibit quantum tunneling selection rules that reflect the intrinsic symmetry of the molecule [45]. With this in mind, future efforts should be directed towards the implementation of magnetic ions possessing stronger spin-orbit anisotropy. Indeed, early studies of rare-earth-based mononuclear species are very promising [76].

## 4. Acknowledgements

This work was supported by the National Science Foundation, grant numbers DMR0804408 and CHE0924374 (SH), DMR0506946 (SH & DNH), CHE0714488 (DNH) and DMR0747587 (EdB). The National High Magnetic Field Laboratory is supported by the NSF (Cooperative Agreement No. DMR0654118) and by the State of Florida.

**Table 1.** Comparisons between $D_{mol}$ and $d_{ion}$. The $D_S$ correspond to molecular axial anisotropy, with the subscript representing the spin state, $S$. The $U_S$ correspond to magnetization barriers, making the further assumption that $U_S = D_S S^2$. Scheme 1 and 2 refer to two possible coupling schemes for the antiferromagnetic Mn$_6$ molecule [16,17].

| Molecule(s) | Relation between $D_{mol}$ and $d_{ion}$ | | $D_S$ and $U_S$ ratios |
|---|---|---|---|
| Mn$_3^{III}$ | $D_6 = {}^3/_{11} d_{ion}$ | | $D_2/D_6 = {}^{253}/_{49} = 5.16$ |
| | $D_2 = {}^{69}/_{49} d_{ion}$ | | $U_6/U_2 = 1.74$ |
| Mn$_6^{III}$ | $D_{12} = {}^3/_{23} d_{ion}$ | | |
| | $D_4 = {}^{21}/_{25} d_{ion}$ | (Scheme 1) | $D_4/D_{12} = 5.4 - 6.4$ |
| | $D_4 = {}^{2967}/_{4235} d_{ion}$ | (Scheme 2) | $U_{12}/U_4 = 1.44 - 1.71$ |
| Mn$_3^{III}$ / Mn$_6^{III}$ | | | $D_6/D_{12} = 2.09$ |
| | | | $U_{12}/U_6 = 1.91$ |

**Table 2.** Comparisons between published magnetic and EPR data obtained for a selection of Mn$_3$ and Mn$_6$ complexes. $J$ values were estimated from dc susceptibility, $S$ values from dc susceptibility and magnetization, $U_{eff}$ from Arrhenius analysis of ac susceptibility, $D_{mag}$ from magnetization fits to Eqn. (1), and $D_{EPR}$ from high-field EPR. Blank cells imply that no data are available. Note: (i) we use the $J\hat{S}_1 \cdot \hat{S}_2$ convention [Eqn. (2)] rather than the $-2J\hat{S}_1 \cdot \hat{S}_2$ convention found in many chemistry journals; and (ii) the numbering of complexes inevitably differs from the schemes employed in the cited references.

| | # | Complex | Space group | $S$ | $J$ (cm$^{-1}$) | $U_{eff}$ (K) | $D_{mag}$ (cm$^{-1}$) | $D_{EPR}$ (cm$^{-1}$) | Refs |
|---|---|---|---|---|---|---|---|---|---|
| Mn$_3$ | 1 | [Mn$_3$O(sao)$_3$(O$_2$CPh)(H$_2$O)(py)$_3$] | P-3 | 2 | | | −2.39 | ∼−7[b] | 19 |
| | 2 | [Mn$_3$O(sao)$_3$(O$_2$C-napth)(py)$_3$]·py | P-1 | 2 | +6.0[a] | | −2.33 | | 19 |
| | 3 | [Mn$_3$O(tBu-sao)$_3$(O$_2$CPh(OMe))$_3$(py)$_3$] | C2/c | 2 | +7.8 | | −3.77 | | 19 |
| | 4 | [NEt$_4$]$_3$[Mn$_3$Zn$_2$(salox)$_3$O(N$_3$)$_8$] | P2$_1$/n | 2 | +7.0[a] | 9.7 | | ∼−5[b] | 25,63 |
| | 5 | [Mn$_3$O(Et-sao)$_3$(O$_2$CPh(Cl)$_2$)(H$_2$O)(MeOH)$_3$] | P-1 | 6 | −3.6 | 43.7 | −0.59 | | 19 |
| | 6 | [Mn$_3$O(Et-sao)$_3$(MeOH)$_3$(ClO$_4$)] | R-3 | 6 | −5.6 | 57.0 | −0.77 | −0.68 | 19 |
| | 7 | [Mn$_3$O(Et-sao)$_3$(Et-py)$_3$(ClO$_4$)] | R-3 | 6 | −8.2 | 48.0 | −0.48 | −0.69 | 19,64 |
| | 8 | [NEt$_4$]$_3$[Mn$_3$Zn$_2$(salox)$_3$O(N$_3$)$_6$Cl$_2$] | R3c | 6 | −3.4 | 44.0 | −0.83[d] | −0.80 | 22,25 |
| | 9 | [NEt$_4$]$_3$[Mn$_3$Zn$_2$(salox)$_3$O(N$_3$)$_6$Br$_2$] | R3c | 6 | −3.2 | 43.7 | −0.77[d] | −0.82 | 22,25 |
| | 10 | [NEt$_4$]$_3$[Mn$_3$Zn$_2$(Me-salox)$_3$O(N$_3$)$_6$Cl$_2$] | R3c | 6 | −6.6 | 45.6 | −0.76[d] | −0.81 | 25 |
| Mn$_6$ | 11 | [Mn$_6$O$_2$(Me-sao)$_6$(O$_2$CCPh$_3$)$_2$(EtOH)$_4$] | C2/c | 4 | +3.8[c] | 31.7 | | −1.27 | 16,18 |
| | 12 | [Mn$_6$O$_2$(sao)$_6$(O$_2$CH)$_2$(MeOH)$_4$] | P-1 | 4 | +9.2[c] | 28.0 | −1.4 | | 18 |
| | 13 | [Mn$_6$O$_2$(sao)$_6$(O$_2$CPh)$_2$(MeCN)$_2$(H$_2$O)$_2$] | P-1 | 4 | | 23.8 | −1.6 | | 18 |
| | 14 | [Mn$_6$O$_2$(sao)$_6$(1-Me-cyclohex)$_2$(MeOH)$_4$] | P-1 | 4 | | 28.8 | −1.2 | | 18 |
| | 15 | [Mn$_6$O$_2$(Et-sao)$_6$(O$_2$CPh)$_2$(EtOH)$_4$(H$_2$O)$_2$] | P-1 | 12 | −1.8 | 53.1 | −0.43 | −0.35 | 16,64 |
| | 16 | [Mn$_6$O$_2$(Et-sao)$_6$(O$_2$CPh(Me))$_2$(EtOH)$_6$] | P2$_1$/n | 12 | −3.2 | 86.4 | −0.43 | −0.36 | 16,18 |
| | 17 | [Mn$_6$O$_2$(Et-sao)$_6$(O$_2$C$_{11}$H$_{15}$)$_2$(EtOH)$_6$] | P-1 | 12 | −3.2 | 79.9 | −0.43 | | 18 |

[a]Fits to a two $J$ parameter model indicate one of the couplings may be very weak (∼0), thereby relieving much of the frustration. [b]EPR data could not be fit to a giant spin model [Eqn. (1)], so these numbers serve only to indicate that the ZFS appears to be very substantial for these complexes. [c]We only quote the antiferromagnetic coupling constant within the triangular Mn$_3$ units here; the ferromagnetic coupling is of the order −1.2 cm$^{-1}$ for both examples. [d]These $D_{mag}$ values were obtained via fits to a multi-spin Hamiltonian [Eqn. (2)]—see Section 2.3(b) and Ref. [25].

**Figure Captions**

**Figure 1.** Schematic structures of (a) $Mn_3^{III}$ and (b) $Mn_6^{III}$: Mn – purple; O – brown; N – blue. Only the bridging atoms that participate in the exchange have been included, i.e. most of the ligands have been omitted.

**Figure 2.** Main panel: plot of $D_4/D_{12}$ as a function of the ratio $J/d_{ion}$, obtained via numerical diagonalization of Eqn. (2) using the sparse matrix algorithm [66]. Inset: separate plots of $D_{12}$ and $D_4$ versus $J/d_{ion}$. The values of $D_{mol}$ were obtained by first computing the low-energy zero-field spectrum, then measuring the energy splitting, $\Delta_0$, between the lowest two energy levels associated with the ground spin multiplet and equating this with $(2S-1)D_S$. We note that such a procedure introduces a small error as $J/d_{ion}$ decreases. This is associated with the emergence of higher-order molecular ZFS [e.g. $B_4^0$, see Section 2.3(c)] in Eqn. (1), which was neglected here. We stress again that our goal here is simply to demonstrate that the correspondence between theory and experiment improves when considering realistic exchange coupling. We do not expect precise agreement given the simplifying assumptions employed. The $J/d_{ion}$ range is limited to two orders of magnitude because the calculations were time consuming.

**Figure 3.** (a) High temperature EPR spectra of [Ni(hmp)(dmb)Cl]$_4$ with the field parallel to the magnetic easy axis (from [14,15]). All eight ($2S$) $\Delta m_S = 1$ EPR transitions within the $S = 4$ ground state are observed, as indicated by the vertical dashed lines; the resonances marked by asterisks correspond to transitions within low-lying excited spin multiplets. For a purely parabolic $m_S$ dependent spectrum, the spacing between peaks should be $\Delta B = 2D/g\mu_B$. As can be seen, the spacing between the peaks is very clearly dependent on $m_S^2$ (see horizontal colored arrows). This represents direct evidence for 4$^{th}$ and higher order molecular ZFS terms. (b) Energies of the lowest lying $m_S$ states of the spin $S = 6$ Mn$_3$ complex **8** as a function of the longitudinal field ($B_L$). The colored data correspond to derivatives of magnetization curves ($dM/dH$) obtained at different temperatures. The shifts of the resonances ($dM/dH$ peaks) to lower field with increasing temperature is indicative of the presence of higher order axial anisotropy terms in the Hamiltonian [45].

**Figure 4.** Mapping between the giant spin [Eqn. (1)] and multi-spin [Eqn. (2)] phenomenologies: dependence of (a) $B_4^0$, (b) $D_6$ and (c) $U_{EPR}$ on the microscopic parameters $J$, $d_{ion}$ and $\alpha$ for a Mn$_3$ triangle. Details of these calculations are given in the main text: the lighter colors assume parallel single-ion ZFS tensors (parallel z-axes, or $\alpha = 0$), while the darker colors include a small tilting ($\alpha = 8.5°$); note also that both $D_6$ and $B_4^0$ are negative. In (a), the dependence of $-B_4^0$ on the ratio of $J/d_{ion}$ is a power law with an exponent of $-1$. Experimental $B_4^0$ data for complexes **6-10** have also been superimposed on the theoretical curves in (a) [19,25,64]. The inset to (c) depicts the effect of a finite $J$ on the parabolic spectrum: light blue – parabolic; red – non-parabolic.

**Figure 5.** Power law dependences of the coefficients, $B_k^q$ ($k$ = order of operator, $q$ = order of rotational symmetry about the axial direction), associated with higher-order Stevens operators obtained via a mapping of the multi-spin phenomenology [Eqn. (2)] onto the giant-spin approximation [Eqn. (1)] for Mn$_3$, assuming $d_{ion}$ = –3.38 cm$^{-1}$. Note that $B_{10}^0$ is negative; the other two coefficients are positive.

**Figure 6.** Tunnel splittings involving the metastable $m_S$ = +5 level corresponding to resonances $k$ = 0, $k$ = 1, $k$ = 2 and $k$ = 3, as a function of the transverse field, $B_T$, with the Jahn-Teller axes aligned along the $z$-axis (thin lines) and tilted $\alpha$ = 8.5° away from the $z$-axis (thick lines). The grey shaded region in the vicinity of zero transverse field represents the approximate strength of the electron dipolar magnetic fields (~250 G) felt by the Mn$_3$ molecules within the single crystal.

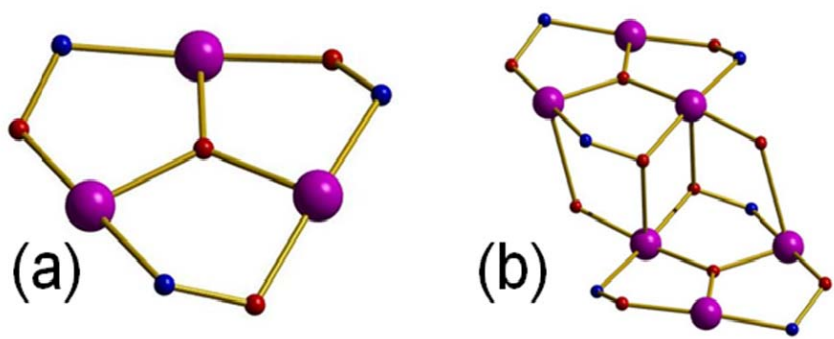

Figure 1. Hill et al.

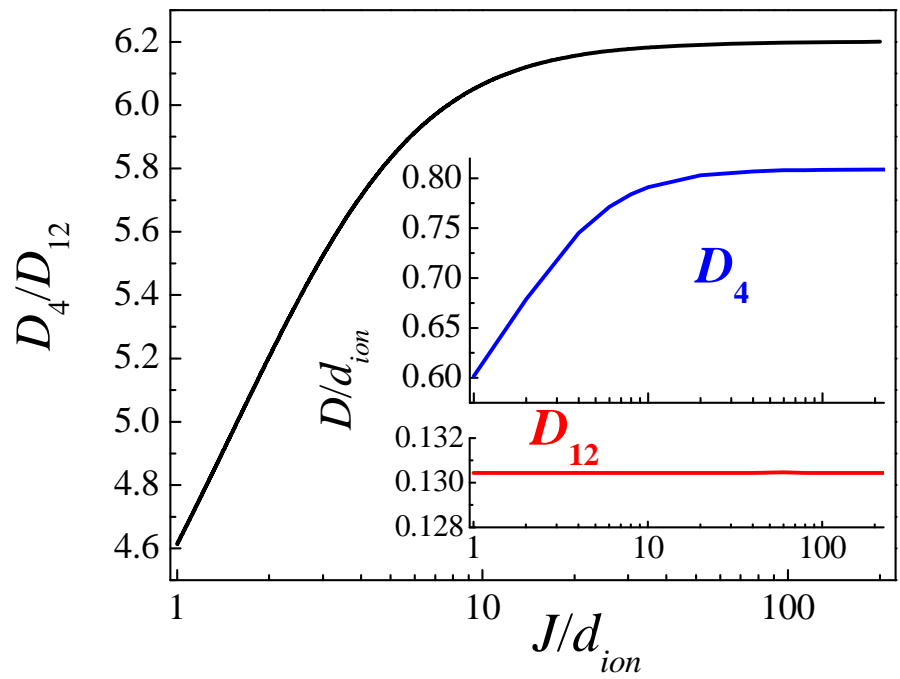

Figure 2. Hill et al.

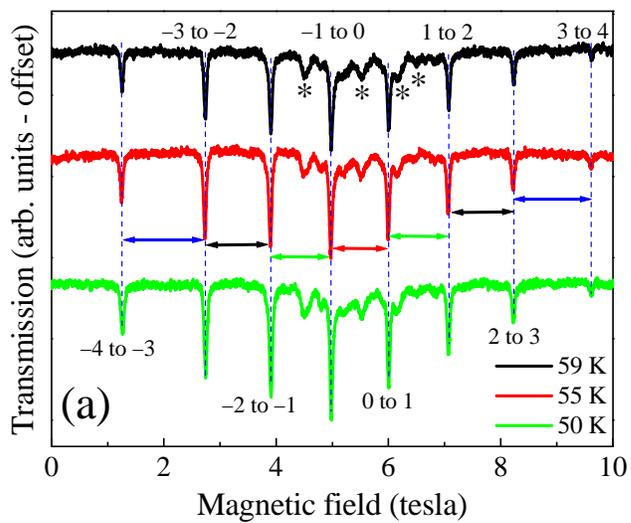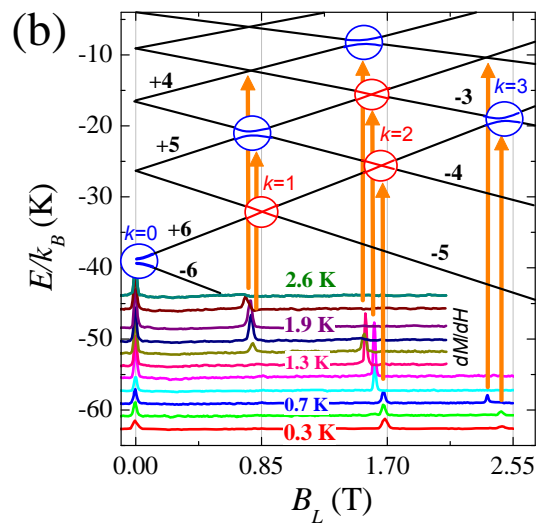

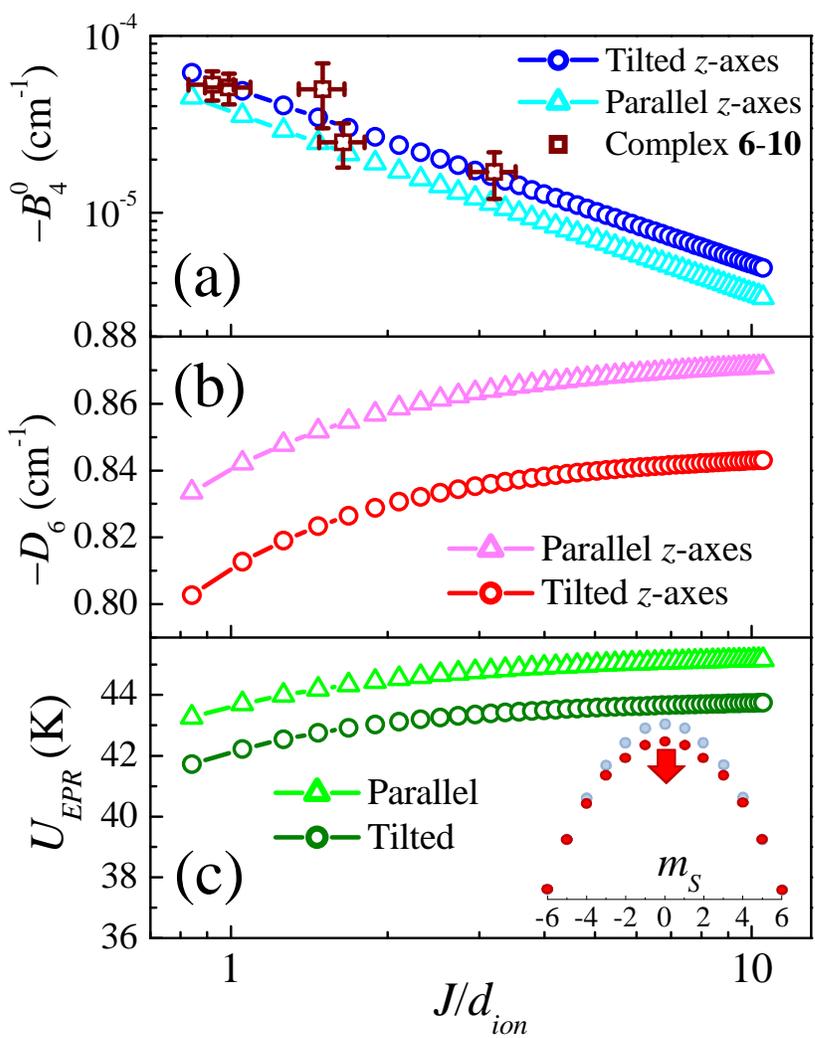

Figure 4. Hill et al.

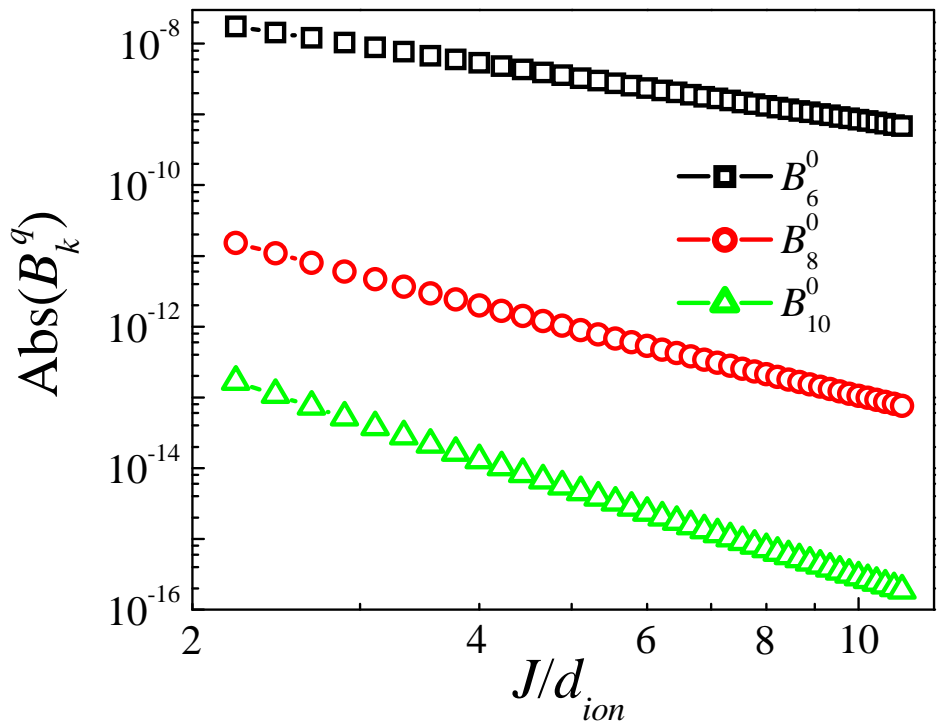

Figure 5. Hill et al.

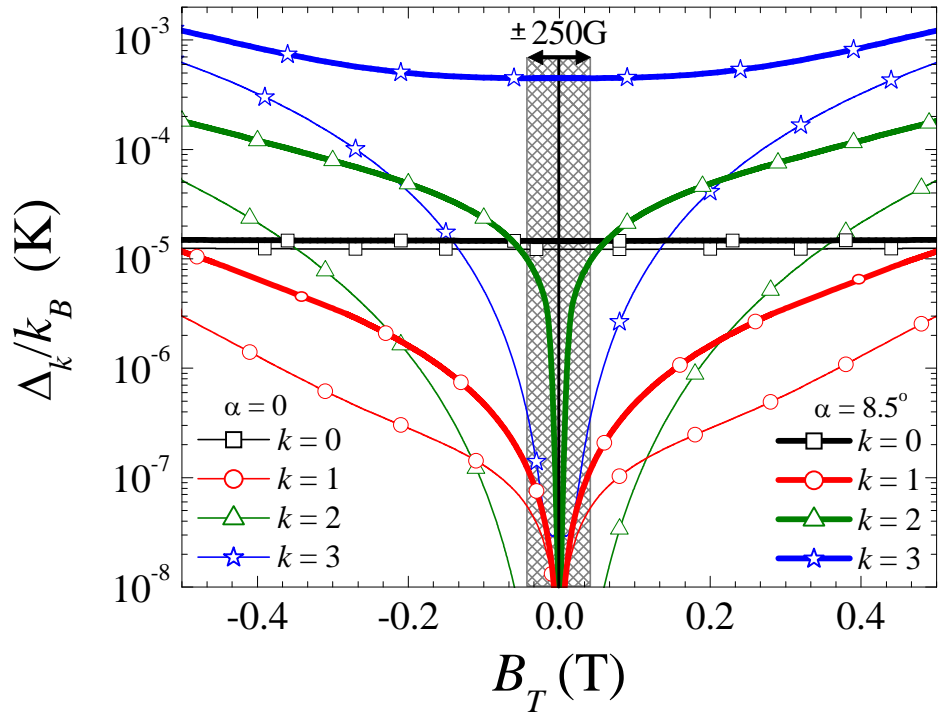

Figure 6. Hill et al.